\documentclass[preprint,showpacs,amsmath,amssymb]{revtex4}
\draft
\usepackage{graphicx,color}

\begin{document}

\title{ Distortion of Wigner molecules 
: pair function approach
}
\author{ M.Taut}
\affiliation{ Leibniz Institute for Solid State and Materials Research, 
IFW Dresden\\
POB 270116, 01171 Dresden, Germany,\\
email: m.taut@ifw-dresden.de}

\date{\today}

\begin{abstract}
We considered  a two dimensional 
three electron quantum dot in a magnetic field in the Wigner limit.
A unitary coordinate transformation 
decouples the Hamiltonian (with Coulomb interaction
between the electrons included) 
into a sum of three independent  pair Hamiltonians.
The eigen-solutions of the pair Hamiltonian provide a spectrum of pair states.
Each pair state defines the distance of the two electrons 
involved in this state. 
In the ground state for given pair angular momentum $m$, 
this distance increases with increasing $|m|$.
The pair states have to be occupied under consideration of the 
Pauli exclusion principle, which  
differs from that for one-electron states and depends on the 
total spin $S$ and the  total  orbital angular momentum $M_L=\sum m_i$ 
(sum over all pair angular momenta).
We have shown that the three electrons
in the ground state of the Wigner molecule 
form an equilateral triangle (as might be expected) only, 
if the state is a  quartet ($S=3/2$) and the orbital angular momentum 
is a magic quantum number ($M_L=3\; m ; m=$ integer). Otherwise
the triangle in the ground state is isosceles.
For $M_L=3 m+1$ one of the sides is 
longer and for $M_L=3 m-1$ 
one of the sides is shorter than the other two.
\end{abstract}

\maketitle

\large

\newpage
\section{Introduction}
Quantum dots are artificial atoms or molecules,
where the electron number, 
the scalar potential and the magnetic field are tunable 
and can provide favorable conditions for all kinds of 
fascinating effects (for recent reviews see 
\cite{Maksym-00,Reimann-02,Yannouleas-07}).
One effect is the formation of  Wigner molecules (WMs). 
 In \cite{Taut-2einB} it has been  considered and 
quantitatively described 
 in the simplest system, the two electron dot, 
using the exact analytical solutions for this system \cite{Taut-2einB}.
In the present paper, another effect, namely a Jahn-Teller-like  distortion of the WM, 
is analyzed. 
Again analytical solutions for the simplest system, where this effect can occur 
(the three electron dot), proved useful. 

From the very beginning 
we should be aware of the fact, that 
Wigner localization
in circular symmetric systems 
cannot be identified using 
the electron density $n({\bf r})$. 
Because the Hamiltonian commutes with the total angular momentum operator, 
the electron density can always be chosen circular symmetric 
\cite{Hirose} and exhibits only a radial shell structure.
Non-circular solutions are an indication of degeneracy, 
but not of Wigner localization.
 The fact that the electrons  keep more or less  
fixed distances from each other due to 
a strong electron-electron correlation 
can be observed in 
the pair correlation function  \\
$G({\bf r})=<\psi|\sum\limits_{i<j}\delta({\bf r}_i-{\bf r}_j-{\bf r})|\psi>$\\
or the two-particle density matrix  \\
$P({\bf r},{\bf r'})=<\psi|\sum\limits_{i<j}\delta({\bf r}_i-{\bf r})
\delta({\bf r}_j-{\bf r'})|\psi>$\\
or in spin resolved versions thereof.
For few electron dots both quantities are equivalent in exhibiting 
the existence of Wigner molecules.
(In \cite{Taut-2einB} the pair correlation function $G({\bf r})$
in conjunction with the density $n({\bf r})$ 
was used for describing the effect
instead of the
nowadays favored two-particle density matrix $P({\bf r},{\bf r'})$).

A useful and illustrative
notion of a WM  in a circular symmetric confinement 
is a rotating  and vibrating finite electron lattice 
\cite{Bolton-93,Maksym-96,Yannouleas-02,Yannouleas-03}.
Due to this picture, a three-electron WM in a circular environment 
 would form a equilateral triangle. 
In the present paper, however, we have shown that this 
is only the case in the classical limit 
where the  distances between the electrons are so large that 
the exponentially decaying overlap between the localized 
 wave functions 
of the individual electrons do not matter, but the 
long range Coulomb interaction is still effective.
Shorter distances between the electrons and consequently 
 overlap between the localized electron wave functions {\em  may} lead to a 
distortion. Whether a distortion occurs depends on the angular momentum. 
This distortion of the seemingly natural  equilateral  symmetry 
is reminiscent of  the Jahn-Teller effect, 
although in the present  case the displaced objects are not atoms,
 but localized single electrons.
(For a recent book on the Jahn-Teller effect including  a lot of references
 see e.g. \cite{Bersuker}.)
As shown below, the ground state of the WM for 
three electrons is isosceles including the 
equilateral as a special case. Excited states can be completely non-symmetric.

Meanwhile there are a couple of papers 
\cite{Laughlin-3e,Hawrylak-93,Pfannkuche-98,Taut-3einB,Mikhailov-02}
which are focused on  the case of three electrons. 
More work, which includes three electrons as a special case,  can be found 
in recent reviews \cite{Maksym-00,Reimann-02}.
The issue of the deformation of the WM without magnetic field
has been discussed   
in \cite{Mikhailov-02} by investigating 
the two-particle density matrix $P({\bf r},{\bf r'})$ using wave functions
 from exact diagonalization in an oscillator eigenfunction basis.
The author found that with increasing coupling parameter 
$\lambda=l_0/a_B$
 ($l_0=\sqrt{\hbar/(m^* \omega_0)}$, $a_B=\hbar^2/(m^* e^2)$)
 there is a level crossing at $\lambda=4.343$, where the ground state  
switches from the $(L,S)=(1,1/2)$ to the
 $(0,3/2)$ state.
He developed the following qualitative picture. 
In the {\em quartet} state 
(ground state for large $\lambda$), the WM forms a 
equidistant triangle for all $\lambda$, whereas 
in the {\em doublet} state (ground state for small $\lambda$)
the spatial distribution is less trivial. Here, 
for infinite $\lambda$ the electrons occupy the edges of an
equidistant triangle. With decreasing $\lambda$
the triangle is increasingly deformed into a isosceles one.
This picture is a special case of the theory presented here.
The purpose of the present paper is to   
extend the considerations to finite $B$ and to present a simple 
quantitative model.

\newpage

\section{Decoupling into three pair problems}
First, we want to review that part of the decoupling of the 
three-electron Hamiltonian 
into three pair Hamiltonians \cite{Taut-3einB}, which is vital
for the general understanding of the current paper.
A numerical check of the validity of this decoupling procedure 
for the three-electron problem in a three-dimensional confinement 
without magnetic field can be found in \cite{Taut-3e-3D}. 
Unlike in \cite{Taut-3einB}, in the present paper we did not 
take advantage of the fact that for certain external field strength and 
quantum numbers there are analytical solutions of the pair equation, 
but instead we solved the radial pair equation 
(one-dimensional eigenvalue problem) numerically, 
whenever a concrete solution is required.
The basic results, however, can be understood without having concrete 
numerical wave functions, but using only the Pauli principle.

The Hamiltonian for three electrons in a homogeneous magnetic field $\bf B$
with the vector potential ${\bf A}({\bf r})=(1/2)\;{\bf B}\times {\bf r}$ 
and a harmonic scalar confinement
(oscillator frequency $\omega_0$) reads
\begin{equation}
H=\sum_{i=1}^3\biggl[{1\over 2}\biggl(\frac{1}{i}{\bf \nabla}_i
+{1\over c}{\bf
A}({\bf r}_i)\biggr)^2 +{1\over 2}\omega_o^2\;r_i^2\biggr]
+\sum_{i<k}{1 \over |{\bf r}_i-{\bf r}_k|}+H_{spin}
\label{h-orig}   
\end{equation}
This is in atomic units $\hbar=m=e^2=1$. 
For a model, where $m$ is replaced by 
an effective electron mass $m^*$ and $e^2$ by the effective (screened) charge 
$e^{*2}=e^2/\epsilon$, our results are in  
effective atomic units ($a.u.^*$) defined by $\hbar=m^*=e^{*2}=1$. 
Energies in our figures, which are generally given 
in units of $\omega_0$ or $\omega_c=B/c $ or a combination thereof, are 
therefore independent of the background parameters.
In order to avoid the dependence of the results on any
material dependent parameters,
the Zeeman term
 $H_{spin}=(g^*/2)\;\mu_B^*\; \sum\limits_{i=1}^3 \;{\bf \sigma}_i \;      
\cdot{\bf B}$
with $\mu_B^*=e^*\hbar/2 m^* c$, $\sigma_z=\pm 1$
 is omitted. 
Besides, the Zeeman term has no influence on the 
focus of this paper, namely the
spatial distribution of the electrons for given quantum numbers.
It only shifts the energies and determines, what the quantum numbers 
of ground state are.

Now, we apply a unitary transformation from the original position 
vectors ${\bf r}_i$ to new ones ${\bf x}_i$
\begin{equation}
 \left[ \begin{array} {c} {\bf x}_1\\{\bf x}_2\\{\bf x}_3\end{array} \right]=
\left[ \begin{array} {ccc}1/3&a&b\\b& 1/3&a\\a&b&1/3\end{array}\right]
\left[ \begin{array} {c} {\bf r}_1 \\ {\bf r}_2 \\ {\bf r}_3 \end{array} \right]
\label{trafo} 
\end{equation}
where $a=1/3-1/\sqrt{3}$ and $b=1/3+1/\sqrt{3}$. 
From the inverse transformation we obtain
\begin{eqnarray}
{\bf r}_1-{\bf r}_2=\sqrt{3}\; \biggl({\bf X}-{\bf x}_3\biggr)
 \nonumber \\
{\bf r}_2-{\bf r}_3=\sqrt{3}\; \biggl({\bf X}-{\bf x}_1\biggr)
\label{diff}\\ 
{\bf r}_3-{\bf r}_1=\sqrt{3}\; \biggl({\bf X}-{\bf x}_2\biggr)
\nonumber
\end{eqnarray}
where  ${\bf X} \equiv \frac{1}{3} \sum_{i=1}^3 {\bf x}_i$ 
is the center of mass (c.m.) in the new coordinates, 
which agrees with the c.m. in the original coordinates 
${\bf R} \equiv \frac{1}{3} \sum_{i=1}^3 {\bf r}_i$.
We do not use the Jacobi transformation (see also Appendix C), 
which separates the 
c.m. coordinate from the relative coordinates 
and which breaks the symmetry between the new quasi-particles.
Instead, our transformation retains the symmetry and the 
c.m. is not an independent variable.
Because any unitary transformation leaves the kinetic energy 
and  the harmonic external potential invariant, 
the  Hamiltonian in the new coordinates  reads
\begin{equation}
H=\sum_{i=1}^3\biggl[{1\over 2}\biggl(\frac{1}{i}{\bf \nabla}_i+{1\over c}{\bf
A}({\bf x}_i)\biggr)^2 +{1\over 2} \; \omega_o^2 \; x_i^2
+\frac{1}{\sqrt{3}} \; {1 \over |{\bf x}_i-{\bf X}|} \biggr]
\label{h-trans} 
\end{equation}
While being still exact, (\ref{h-trans}) is not completely
decoupled because ${\bf X}$
contains all coordinates. 
In the Wigner limit, however, where the uncertainty
 of the c.m. vector ${\bf X}$ 
is small compared with the mean  electron- electron distance, we can 
neglect ${\bf X}$ in the denominator of the 
interaction term in (\ref{h-trans}). 
As shown in the Appendices  A and B, the quantitative  errors introduced by 
this approximation 
are small and do not invalidate any of the 
qualitative conclusions of the present paper.
Then, the Hamiltonian in zero order in ${\bf X}$
\begin{equation}
H^{(0)}=\sum_{i=1}^3 h_i
\end{equation}
decouples into a sum of three independent 
{\em pair Hamiltonians}, which can be rewritten as
\begin{equation}
h=-{1\over 2}{\bf \nabla}^2
+{1\over 2} \; \tilde\omega^2 \; x^2 + {1\over 2} \;\omega_c \; l_z
+\frac{1}{\sqrt{3}} \; {1 \over |{\bf x}|}
\label{h-i}   
\end{equation}
Here, $\tilde\omega=\sqrt{\omega_0^2+(\omega_c/2)^2}$ 
is an effective confinement frequency,
$\omega_c=B/c$ is the cyclotron frequency, and $l_z$
is the orbital angular momentum operator.
This suggests the definition of a {\em pair equation}
\begin{equation}
h\; \varphi_{q}({\bf x}) = \varepsilon_{q} \; \varphi_{q}({\bf x})
\label{pair-eq}   
\end{equation}
with the normalization condition
\mbox{$ \int d^2 {\bf x} \; |\varphi_{q}({\bf x})|^2 =1$}.
The subscript $q$ comprises all quantum numbers.
The {\em pairs} are essentially {\em quasi-particles}.

\begin{figure}[h]
\begin{center}
\includegraphics[width=10cm,angle=270]{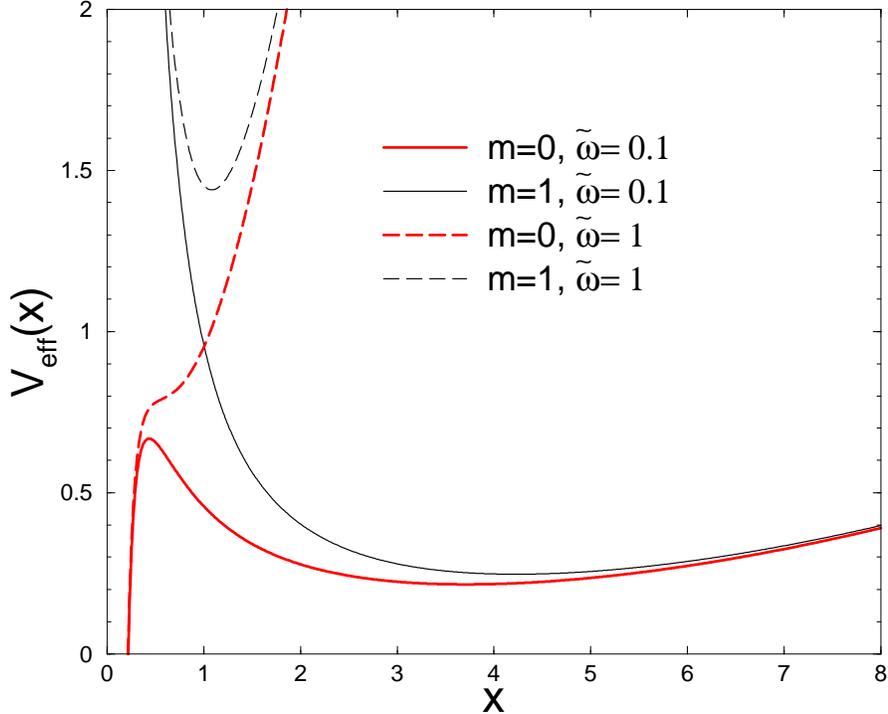}
\caption[]{(Color online) Effective pair potential for two typical effective 
confinement frequencies $\tilde\omega$.}
\label{fig-Veff}  
\end{center}
\end{figure}

In polar coordinates ${\bf x}=(x,\alpha)$ 
we can make the following ansatz for the {\em pair functions}
\begin{equation}
\varphi={e^{im\alpha}\over \sqrt{2\pi}} \; {u(x)\over
x^{1/2}}~~~~~;~~~~~m=0,\pm 1,\pm 2,\ldots
\label{ansatz}   
\end{equation}
where $m$ is an eigenvalue of $l_z$ and therefore 
the angular momentum of a pair.
Inserting (\ref{h-i}) and (\ref{ansatz}) into (\ref{pair-eq}) provides 
 the {\em  radial pair equation}
\begin{equation}
\biggl[-{1\over 2}~{d^2\over dx^2}+V_{eff}(x)
\biggl]u(x)=\tilde\varepsilon \; u(x)
\label{rad-eq}   
\end{equation}
with the {\em effective pair potential}
\begin{equation}
V_{eff}(x)={1\over 2}\biggl(m^2-{1\over 4}\biggr)
{1\over x^2}+{1\over 2}\tilde\omega^2 x^2
+{1 \over \sqrt{3} \; x}
\label{Veff}    
\end{equation}
and the definition 
\begin{equation}
\tilde\varepsilon=\varepsilon-{1\over 2}m \omega_c .
\label{def-epsilon-tilde}  
\end{equation}
The normalization condition reads
\mbox{$\int\limits^\infty_o dx \;|u(x)|^2=1$}.
Fig.\ref{fig-Veff} shows $V_{eff}(x)$ for two typical
 effective confinement frequencies $\tilde\omega$.
Observe that  $V_{eff}(x)$ for $m=0$ has a minimum at non-zero
 $x$ only for 
$\tilde\omega < \tilde\omega_{cr} = \sqrt{3}/2 = 0.866$.

Because of the decoupling in zero order,
the total eigenvalues and orbital eigenfunctions of $H^{(0)}$ read
\begin{eqnarray}
E_{q_1,q_2,q_3}&=&\varepsilon_{q_1}+\varepsilon_{q_2}+\varepsilon_{q_3} 
\label{E-tot}\\  
\Phi_{q_1,q_2,q_3}({\bf x}_1,{\bf x}_2,{\bf x}_3)&=&\varphi_{q_1}({\bf x}_1)
\cdot \varphi_{q_2}({\bf x}_2)\cdot \varphi_{q_3}({\bf x}_3)
\label{WF-tot}  
\end{eqnarray}
The total energy is a sum of pair energies and the total orbital eigenfunction 
is a product of pair functions.
In the original coordinates, the orbital eigenfunctions read
\begin{equation}
\Phi_{q_1,q_2,q_3}({\bf r}_1,{\bf r}_2,{\bf r}_3)=
\phi_{q_1}(2 - 3)\cdot \phi_{q_2}(3 - 1) \;\cdot\; \phi_{q_3}(1 - 2)
\label{WF-rspace}  
\end{equation}
where, for the sake of obtaining simpler formulae, we introduced the following 
shorthand notation
 $\phi(i - k)=\varphi\left( {\bf R}-({\bf r}_i - {\bf r}_k)/\sqrt{3}
\right)$. 

In order to avoid a basic misunderstanding, we want to stress that 
our independent-pair picture does {\em not} imply a division 
of all electrons into pairs, 
where each electron belongs exactly to one pair,
as familiar from geminal approaches in Quantum Chemistry
(see recent reviews in \cite{geminal}) 
and which works only for even electron number.
Instead, the total WF (\ref{WF-rspace}) 
is a product of two-electron (pair) functions, 
where each electron is involved in
pairs with any of the other electrons.
In \cite{Taut-3einB} it has been shown that 
the high field limit of our wavefunction (WF) for ${\bf R}=0$ 
agrees with the Laughlin WF for three electrons 
\cite{Laughlin-3e}.

Now we want to discuss in an illustrative way, 
how the total WF
(\ref{WF-rspace}) reflects the basic property of WMs, namely the 
strong e-e-correlation. 
Let's consider the limit ${\bf R}={\bf X}=0$ first. 
The radial part of the pair functions $u(x)$ shows a peak at a certain 
coordinate value around the minimum
 of $V_{eff}$ (see Fig.\ref{fig-localization}),
 which we call the pair length and which will be calculated 
in Sect. 4.  This means that the probability density $|\phi_{q_1}(2-3)|^2$ 
of pair $q_1$ is peaked in $\bf r$-space 
whenever $|{\bf r}_2-{\bf r}_3|$ equals the pair length.
Considering the special cyclic 
structure of the total WF (\ref{WF-tot}) 
we can conclude:
The probability density has peaks, if 
the three electrons form a triangle with sides equaling the three 
pair length. The angular orientation of this triangle is arbitrary, 
and its translational location is defined by ${\bf R}=0$. 
This triangle is blurred by the finite width of the peak in $u(x)$.

Although our theory is aimed at the strong correlation limit, 
we want to point out that also the weak correlation limit 
 can be described by product wave functions in the new coordinates.
(This will also be demonstrated  in the next Section by investigating the 
 numerical solutions of the pair equation in this limit.) 
The reason for the decoupling of the Hamiltonian (\ref{h-trans}), 
however, is quite different. 
Whereas in the strong correlation limit decoupling occurs because 
$\bf X$ can be neglected versus ${\bf x}_i$, this is not the case 
for weak correlation. 
In the weak correlation limit, however, the whole e-e-interaction 
term can be neglected versus the kinetic 
energy. This leads to decoupling as well. 
Describing non-interacting electrons by products of WF 
in the new coordinates 
rather than simply by one-electron functions is 
just  a complicated (but equivalent) way of describing the same system.
This ambiguity  is a consequence of the fact that 
(after neglecting the e-e-interaction) our Hamiltonian
is invariant under any unitary transformation of the coordinates.

Because the pair functions are eigenfunctions of 
the orbital  angular momentum operator in $\bf x$-space with eigenvalues $m_i$, 
the total orbital eigenfunctions (\ref{WF-tot}) are eigenfunctions 
of the total orbital angular momentum in ${\bf x}$-space
with eigenvalues $M_L=\sum_{i=1}^3 m_i$.
Because the transformation back to the  ${\bf r}$-space is unitary,
(\ref{WF-rspace}) has the same eigenvalues $M_L$ like (\ref{WF-tot}).
Moreover, through (\ref{ansatz}) and (\ref{WF-rspace}), 
$M_L$ is linked to the parity
of the total wave function. Even  (odd) $M_L$ means even (odd) parity, i.e.
symmetry (antisymmetry) with respect to inversion of all coordinate vectors.

\newpage

\section{Pair energies and electron localization}
\begin{figure}[h]
\begin{center}
\includegraphics[width=10cm,angle=270]{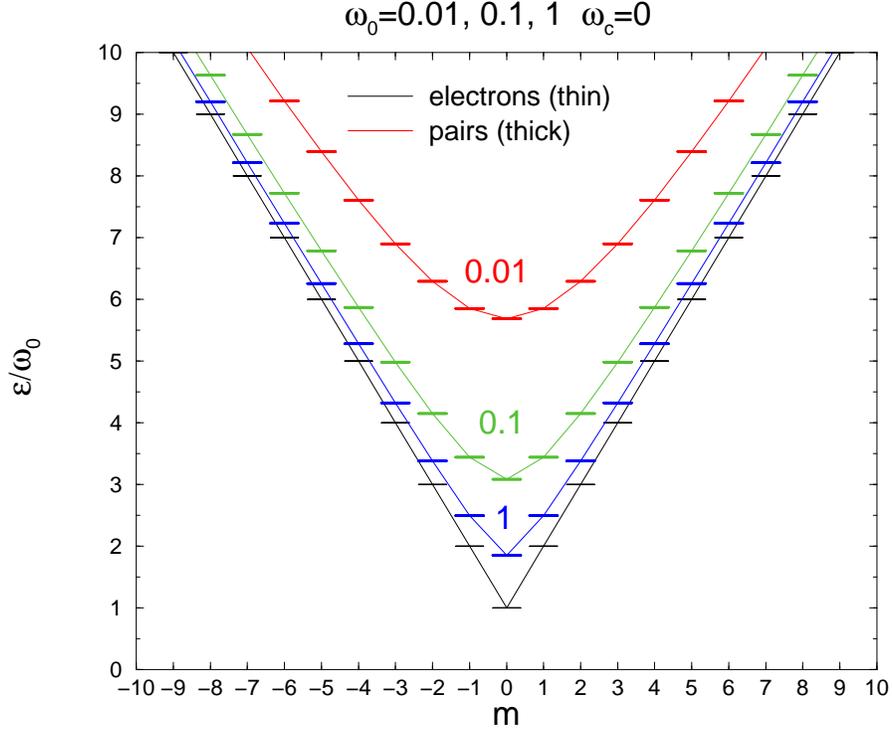}
\caption[]{(Color online) Pair energies (lowest state for given $m$)
for zero magnetic field ($\omega_c=0$) for the three
confinement frequencies indicated on the levels. 
The colored  values are obtained by numerical solution of (\ref{rad-eq}).
The thin (black) levels show the one-particle energies 
for non-interacting electrons given in (\ref{non-int}).
The lines connecting the levels are a guide for the eye.}
\label{fig-E-omega_c=0} 
\end{center}
\end{figure}

\begin{figure}[h]
\begin{center}
\includegraphics[width=10cm,angle=270]{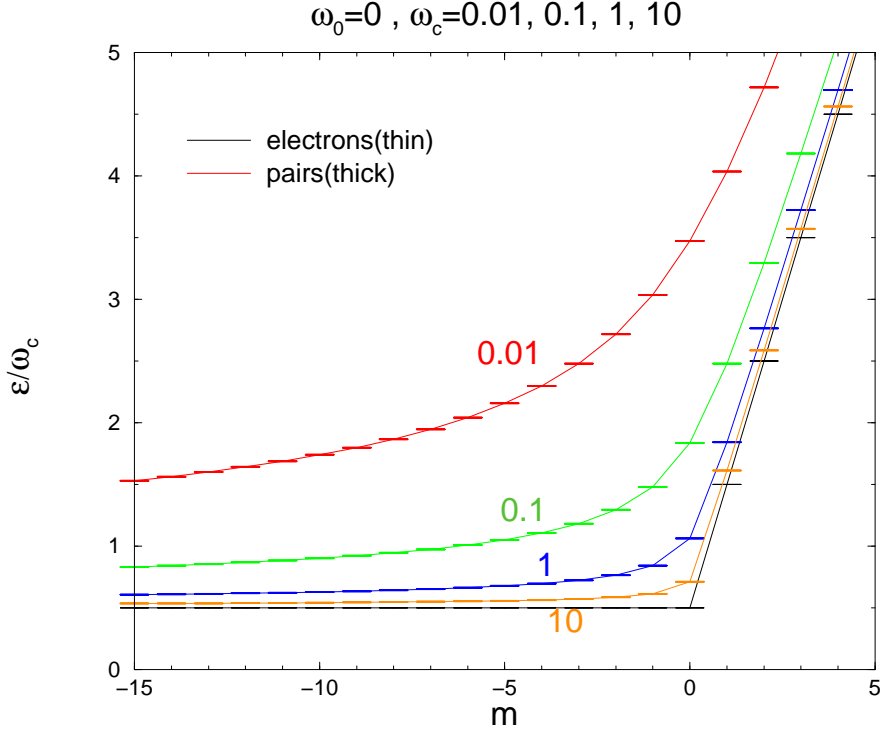}
\caption[]{(Color online) Pair energies (lowest state for given $m$)
without confinement potential
($\omega_0=0$)
for the three cyclotron frequencies indicated on the levels.
The thin (black) levels show the one-particle energies for non-interacting electrons.
}
\label{fig-E-omega_0=0}  
\end{center}
\end{figure}

\begin{figure}[h]
\begin{center}
\includegraphics[width=10cm,angle=270]{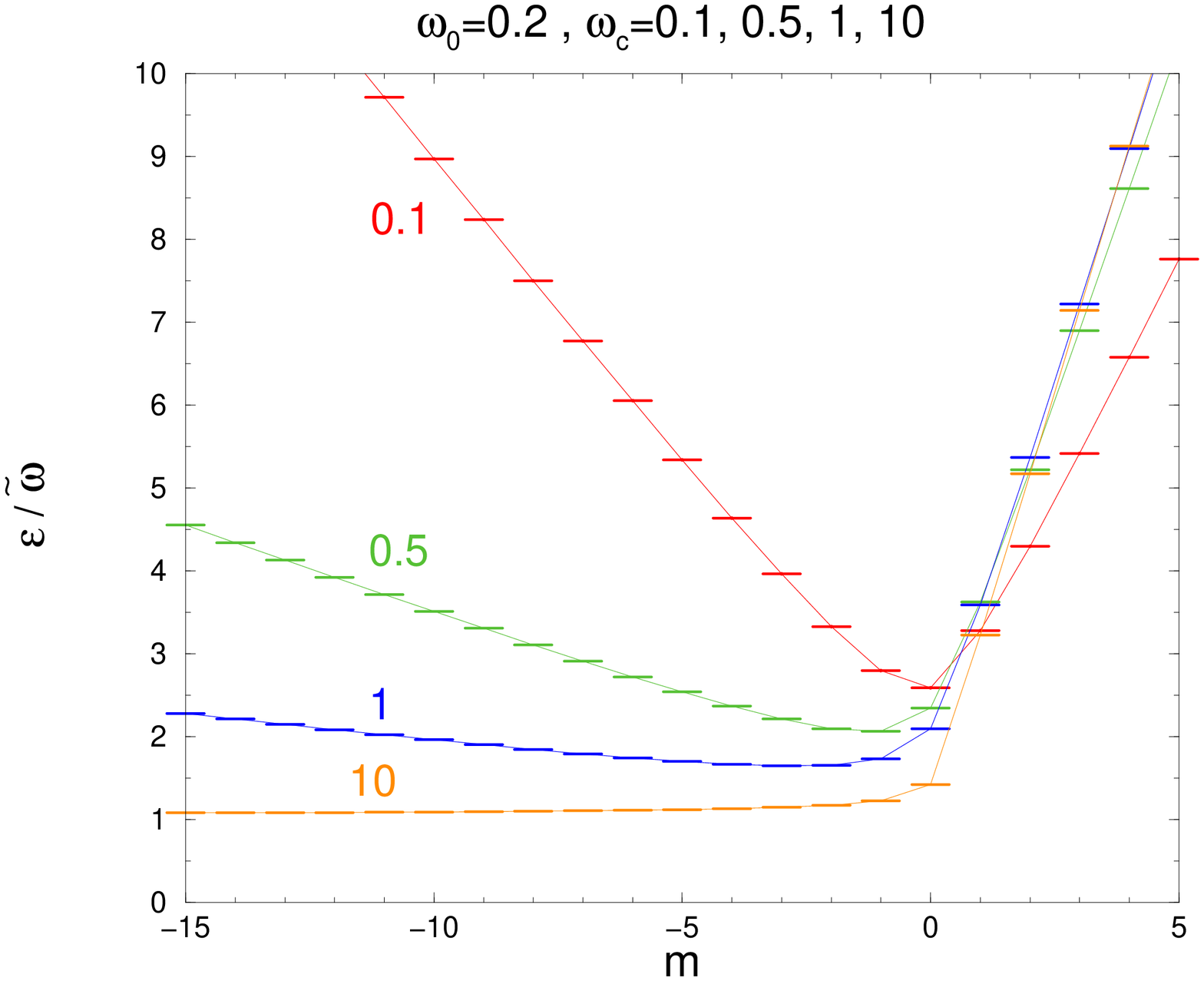}
\caption[]{(Color online) Pair energies (lowest state for given $m$) for
 a dot confinement ($\omega_0=0.2$)
for the four cyclotron frequencies indicated on the levels. }
\label{fig-E-omega_0=.2}   
\end{center}
\end{figure}

The pair energies $\varepsilon_q$ in our pair approach
are of the same central importance for the electronic structure of the system
as the one-particle energies in 
independent particle systems.
The major qualitative difference between both approaches 
lies in the occupation (Pauli principle) of the energy levels, which 
has been investigated in Ref.\cite{Taut-3einB}.
Figs.\ref{fig-E-omega_c=0}-\ref{fig-E-omega_0=.2} show 
the dependence of the ground state energies of the pair levels on the 
orbital angular momentum $m$ of the pairs for fixed external field.\\
Fig.\ref{fig-E-omega_c=0} applies to the limiting case
 of vanishing magnetic field. 
In the limit $\omega_0 \rightarrow \infty$ 
the pair levels converge to the one-electron levels for 
non-interacting electrons (Fock-Darwin levels)
\begin{equation}
\varepsilon_{nm}^{non-int}=(2\; n + |m| + 1) \;
\tilde{\omega}+\frac{1}{2} m\; \omega_c
\label{non-int}   
\end{equation}
where the radial quantum number $n=0,1,2, \cdots$ is the degree of excitation
for given angular momentum  $m$. 
This can be seen in the radial pair equation (\ref{rad-eq},\ref{Veff}):
 For large 
$\tilde{\omega}$ the state is compressed into the region of small $x$, 
where the centrifugal term ($\propto 1/x^2$) is much larger than
 the e-e-interaction term ($\propto 1/x$) and the latter can be neglected. 
Without the e-e-interaction term the pair equation agrees with the one-electron 
Schr\"odinger equation.

Fig.\ref{fig-E-omega_0=0} shows the complementary case of vanishing
confinement. In the limit of strong magnetic fields, 
these pair levels approach the ground states (for given $m$) 
of the Fock-Darwin
levels  (\ref{non-int}).
The analytical verification of this statement is evident in
eqs. (\ref{rad-eq},\ref{Veff}).
It is also clear intuitively, because the strong magnetic field 
 out-plays the effect of the 
Coulomb interaction.\\
A case with a fixed finite confinement $\omega_0$,
which is of the order of magnitude of quantum dots in GaAs 
(in effective atomic units) is shown in Fig.\ref{fig-E-omega_0=.2}.
The angular momentum $m_{min}$ with minimum pair energy is now non-zero.

\begin{figure}[h]
\begin{center}
\includegraphics[width=10cm,angle=270]{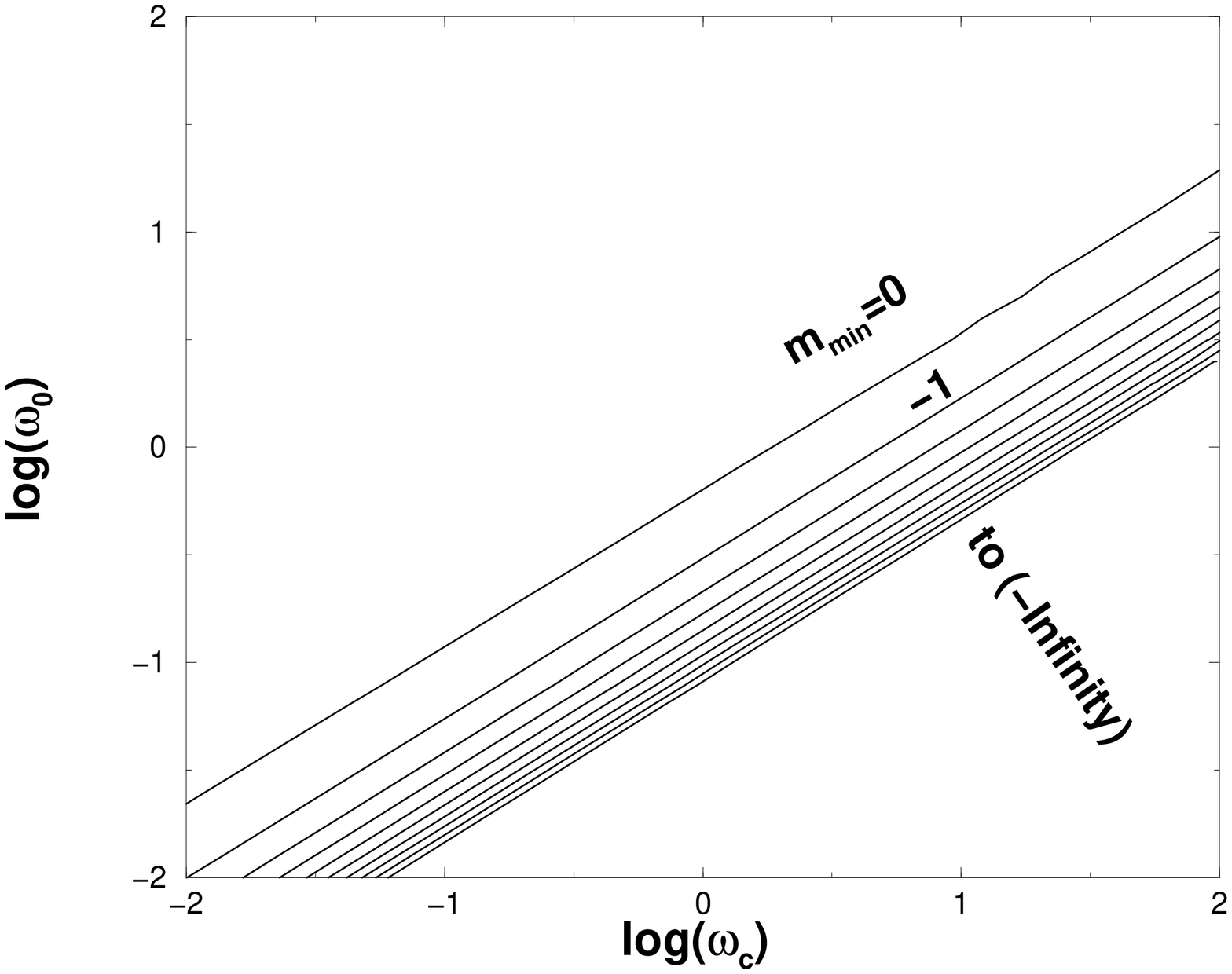}
\caption[]{Orbital angular momentum $m_{min}$ of the  pair
with minimum pair energy for given
cyclotron frequency $\omega_c$ and
confinement frequency $\omega_0$.
 }
\label{fig-phase-diagram}
\end{center}
\end{figure}

\begin{figure}[h]
\begin{center}
\includegraphics[width=10cm,angle=270]{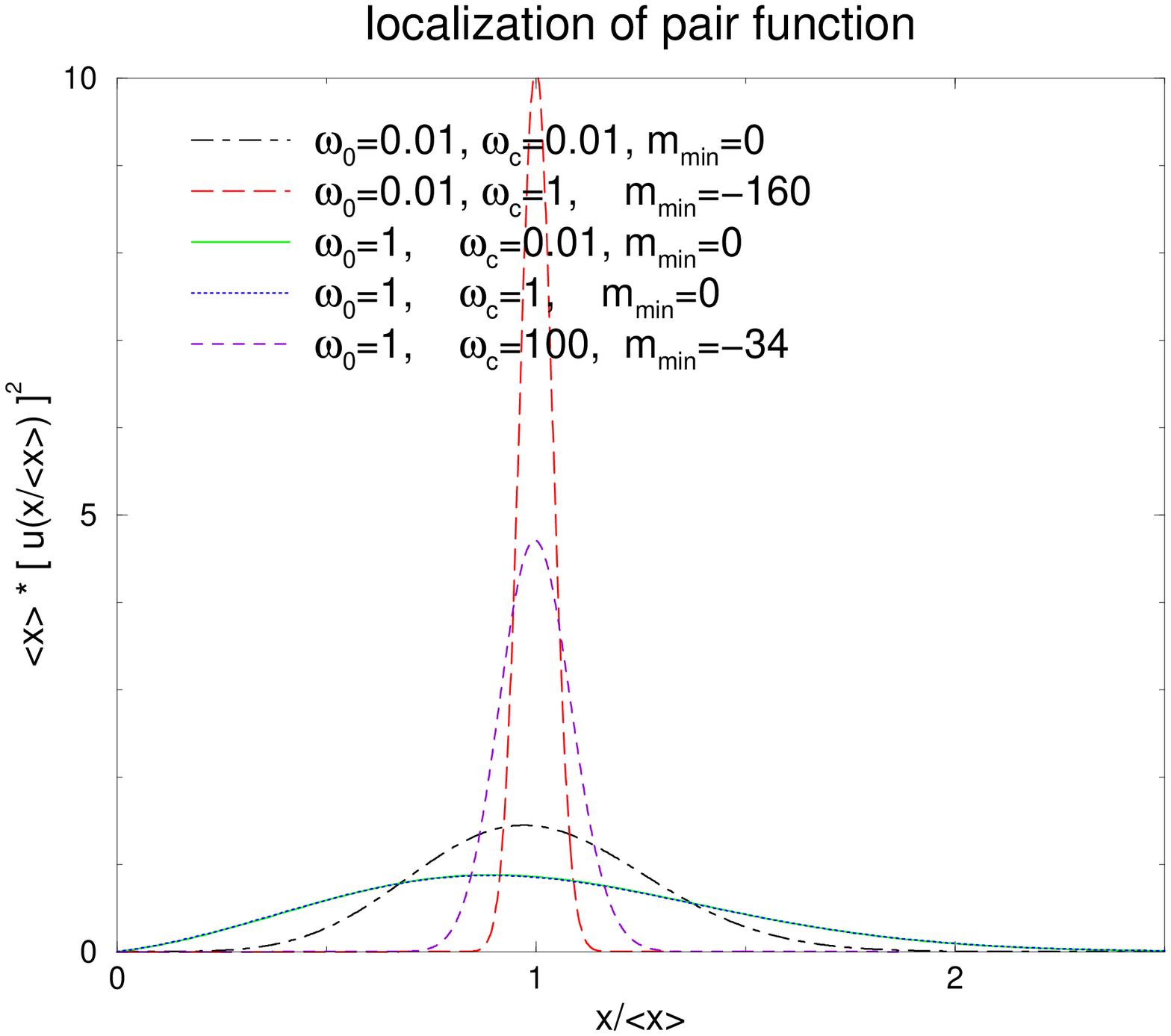}
\caption[]{Localization of the pair function for the external fields 
given in the caption. $m_{min}$ is the angular momentum of the state 
with lowest energy and $<x>$ is the expectation value $<x>=\int dx \;x \;u(x)^2$.
The curves are renormalized to conserve the normalization.
 }
\label{fig-localization}
\end{center}
\end{figure}

Fig.\ref{fig-phase-diagram} provides the angular momemtum $m_{min}$ 
of the lowest pair energy 
for given magnetic field and parabolic scalar confinement.
The lines separate the regions with adjacent $m_{min}$.
In the upper left part of the plot $m_{min} = 0$ and 
and in the lower right part $m_{min}$ converges successively to $-\infty$.
In between only the lines separating the 10 lowest $m_{min}$ 
are shown.
Although only a limited region of the parameter space is shown, the general
features are obvious: All phase boundaries in the log-log-plot 
are well represented by 
parallel straight lines $log(\omega_0)=log(A)+B\cdot log(\omega_c)$.
Consequently, the boundaries in a linear plot
are power functions 
$ \omega_0= A \cdot \omega_c^B$
where $B \approx 0.75$ is universal and the factor $A$ depends on the boundary in question.

Fig.\ref{fig-localization} shows how 
electron localization can be visualized in our approach.
Localization means that (the radial part of) the pair function $u(x)$, 
which describes the electron- 
electron distance, is peaked at a finite x.
There are two reasons for localization:\\
a) for small $\omega_0$ (and small or vanishing $\omega_c$) the
electrons are pushed away from each other by 
 the (last) interaction term in (\ref{Veff}).
On the other hand,\\
b) for large $\omega_c$ (and small $\omega_0$)
 the modulus of the angular momentum of the ground state is large 
(see Fig.\ref{fig-phase-diagram}).
 In this case the separation of the electrons is caused by 
the (first) centrifugal term in (\ref{Veff}).
This mechanism does not work for vanishing $\omega_0$ because 
then the angular momentum is ill defined. It does not work for 
vanishing electron- electron interaction either.\\
Therefore, electron-electron interaction alone can localize the electrons 
without the assistance of the magnetic field, 
but the magnetic field alone cannot do the job. 
As seen in Fig.\ref{fig-localization}, the strongest localization is gained 
for small $\omega_0$(=0.01) and  large (or at least medium) $\omega_c$.
On the other hand, a medium  $\omega_c$(=1)
 cannot achieve anything if 
$\omega_0$ is of comparable size. 
(The curve for $\omega_0$=1, $\omega_c$=0.01 agrees almost 
completely with the curve for $\omega_0$=1, $\omega_c$=1.)
In this case we need a very 
strong magnetic field $\omega_c$(=100) for localizing the electrons.
A quantitative measure for localization is the mean square deviation 
 of $x$ from the expectation value $<x>$. 
$$ (\Delta x)^2=\int_0^\infty dx\;[u(x)]^2\;(x-<x>)^2 $$
The results read in the same order as 
shown in Fig.\ref{fig-localization}: $(\Delta x)^2=$ 0.0749, 0.00155, 0.1987, 0.2017, 0.00717.


\section{Pair length and distortion of the Wigner molecule}

\begin{figure}[h]
\begin{center}
\includegraphics[width=7cm,angle=0]{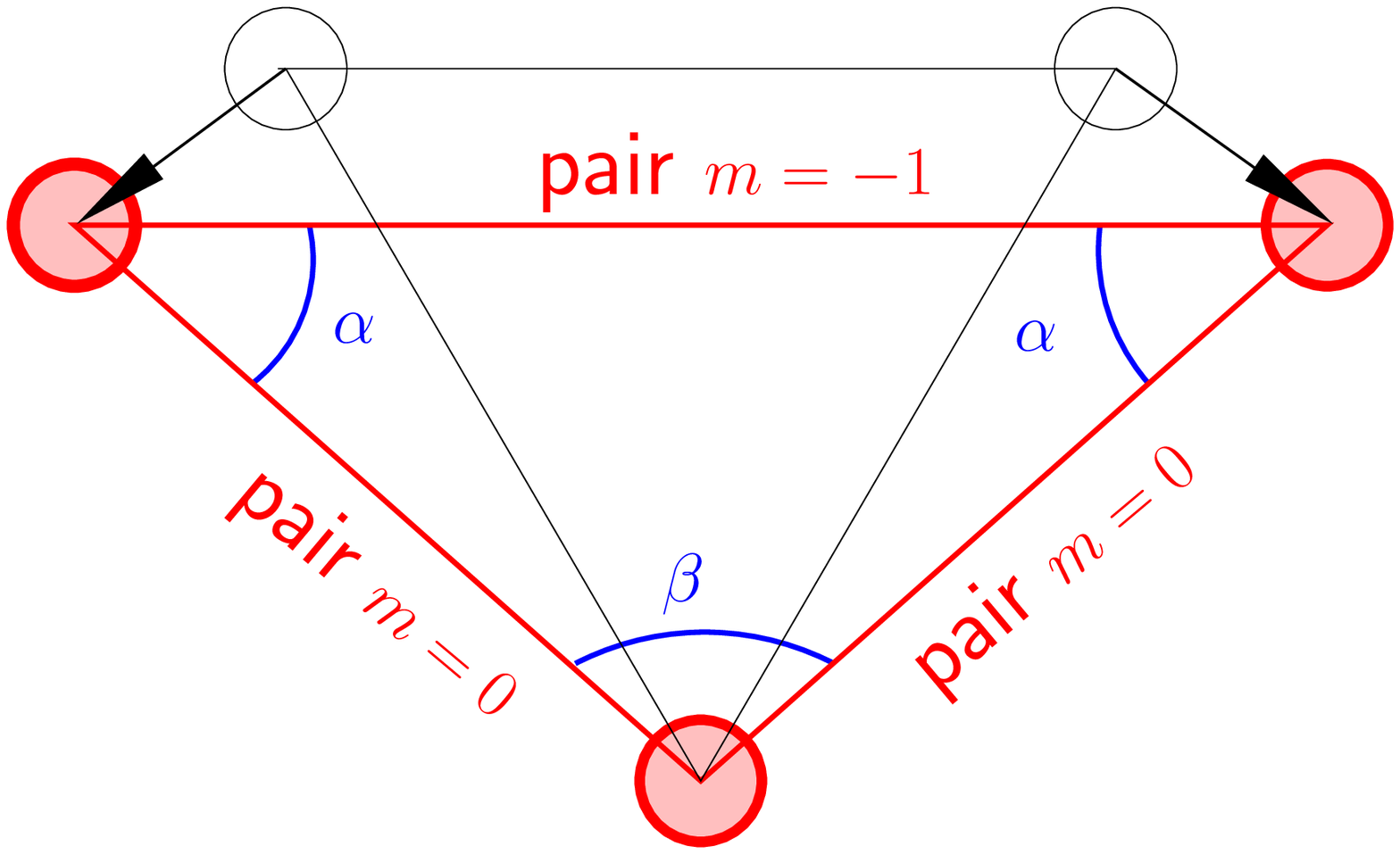}
\hspace{1cm}
\includegraphics[width=5.2cm,angle=0]{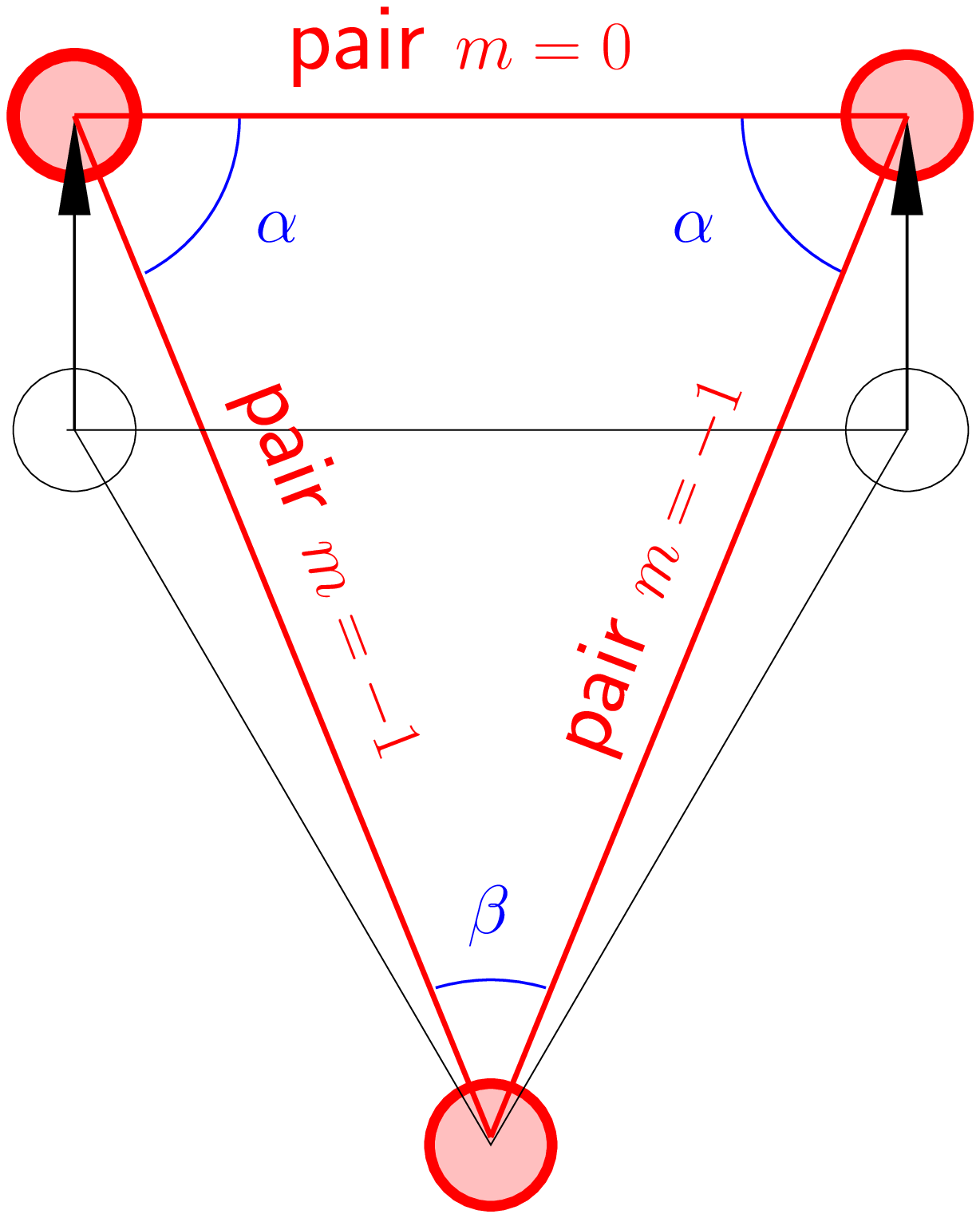}
\caption[]{(Color online) Schematic picture  of  the
distorted three-electron Wigner molecule
with total orbital angular momentum $M_L=-1$ (left) and $M_L=-2$ (right).
The thin lines depict the undistorted WM.}
\label{fig-scetch}
\end{center}
\end{figure}

\begin{figure}[h]
\begin{center}
\includegraphics[width=8cm,angle=270]{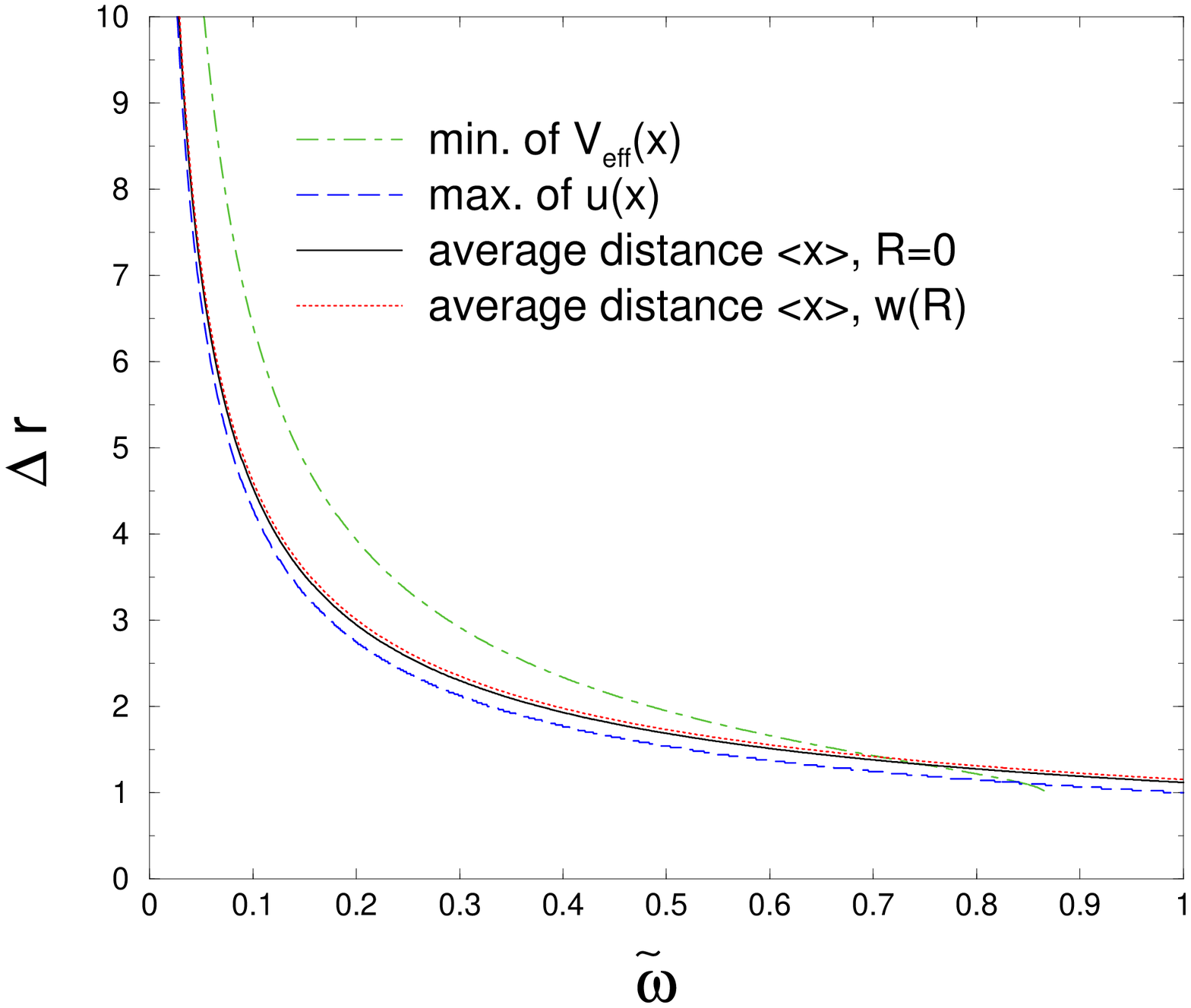}
\caption[]{(Color online) Pair length for $m=0$ as
estimated from different procedures (see text).}
\label{fig-distance-diff-methods} 
\end{center}
\end{figure}

\begin{figure}[h]
\begin{center}
\includegraphics[width=8cm,angle=270]{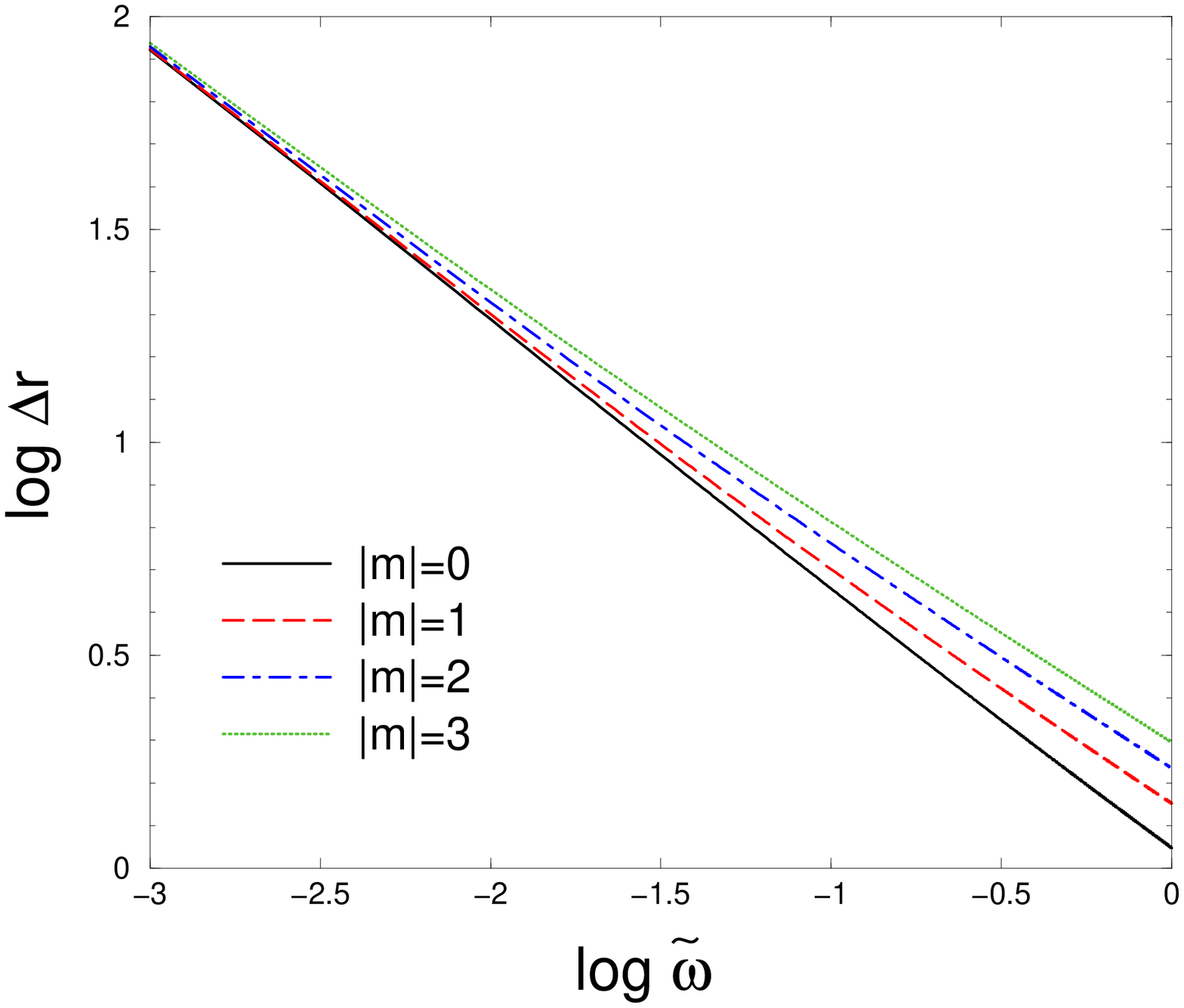}
\caption[]{(Color online) Pair length
calculated from  $<x>$
for several pair
orbital angular momenta $|m|$.}
\label{fig-distance}  
\end{center}
\end{figure}

As seen in Fig.\ref{fig-scetch}, the pair length define
the distortion of the Wigner molecule.
The pair length, or mean electron-electron distance,
can be obtained from the following quantities:\\
(i) the  minimum position of the effective pair potential $V_{eff}(x)$\\
(ii) the maximum position of the radial pair function $u(x)$\\
(iii) the average electron distance
\begin{equation}
<x>=\int d^2{\bf x}\;x\;|\varphi({\bf x})|^2 =\int dx \;x\;[u(x)]^2
\label{x-av}
\end{equation}
The real electron- electron distance $\Delta r$ is 
obtained from the above defined values by multiplication with $\sqrt{3}$ 
as seen from (\ref{diff}).
Definition (iii) agreed with (ii) if the radial pair function was symmetric
with respect to the maximum, and (ii) agreed with (i) if it was
$\delta$-function like.
In our curves discussed below, the
radial wave functions of the {\em ground state }
for the corresponding $m$ have been used throughout.
Observe that  for $m=0$ and $\tilde{\omega} > \sqrt{3}/2$
 the first definition breaks down completely, because $V_{eff}(x)$
has no minimum at  non-zero $x$ (see also Fig.\ref{fig-Veff}).
Fig.\ref{fig-distance-diff-methods} compares these definitions for
$m=0$. As to the agreement of the first and simplest definition
with the two more sophisticated ones
one has to consider, that a discrepancy of this size occurs
only for $m=0$, where the centrifugal potential in 2D is negative.
The difference is much less for $|m| > 0$.
Fig.\ref{fig-distance} shows the dependence of the pair length
 $\Delta r$  on the orbital angular momentum $m$.
Apart from getting an idea about the order of magnitude of the dependence,
we notice that the pair length is the larger the larger $|m|$ is.
(Observe that $V_{eff}(x)$ and $u(x)$ depend only on the modulus of $m$.)
This follows directly from the fact, that the centrifugal potential
in (\ref{Veff}),
which grows with growing $|m|$,
pushes the electrons within a pair away from each other.
This effect can also be understood on the basis of classical mechanics.
From the almost linear behavior in the log-log-scale we conclude, that the 
function of $\Delta r$ versus $\widetilde{\omega}$ is roughly a power law.

\begin{figure}[h]
\begin{center}
\includegraphics[width=8cm,angle=270]{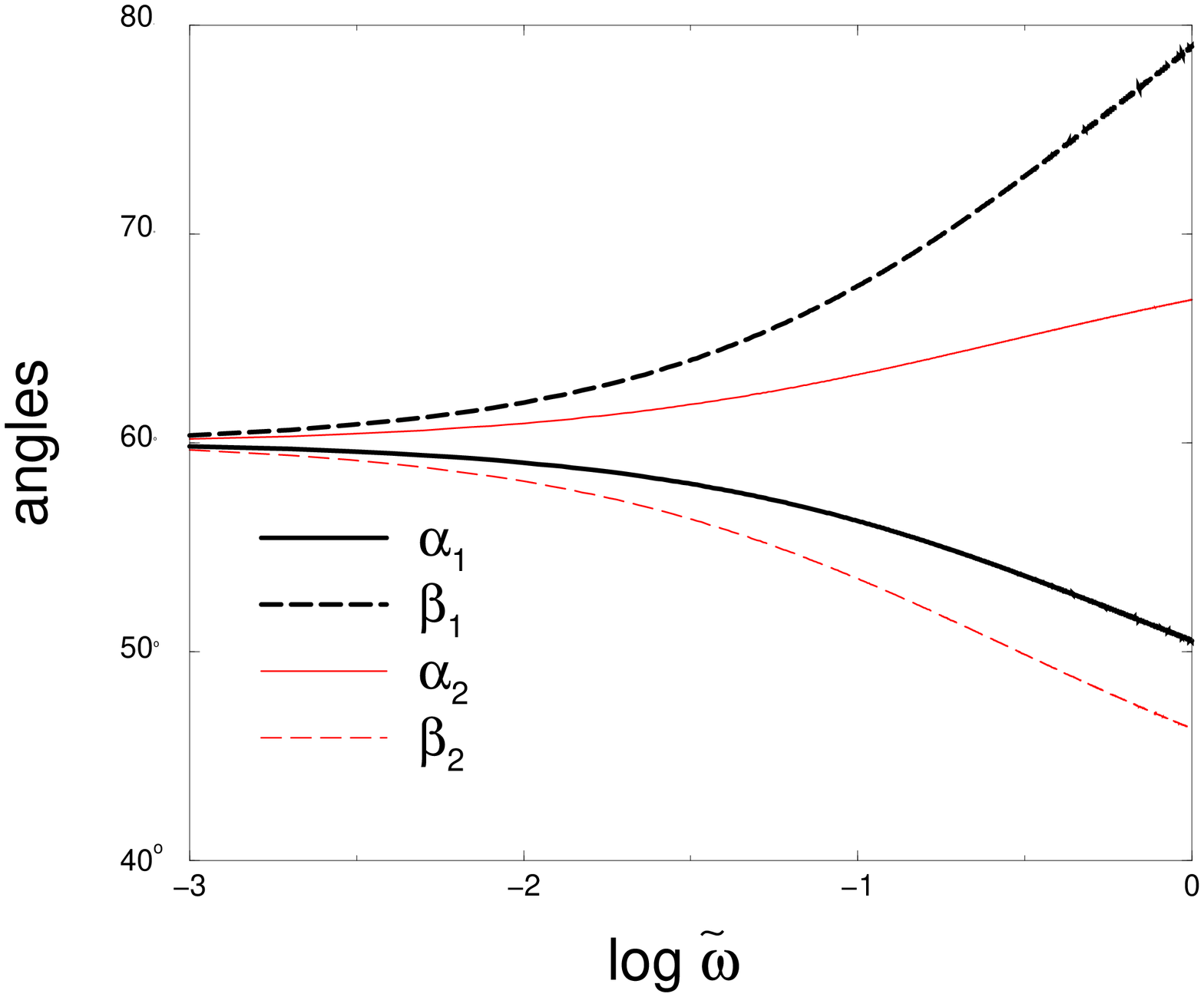}
\caption[]{(color online) Angles in the distorted Wigner molecule
calculated from $<x>$.
The thick (black) lines refer to the case where two
pairs have $m=0$  and one pair has $m=-1$
as sketched in Fig.\ref{fig-scetch} (left).
The thin (red) lines belong to the case where two pairs have $m=-1$
and one pair has $m=0$ as shown in Fig.\ref{fig-scetch} (right). 
As in Fig.\ref{fig-scetch},
$\alpha$ gives the values of the two equal angles and $\beta$
is the third angle.
}
\label{fig-angles}
\end{center}
\end{figure}

For ground states,
the qualitative features of the distortion
may be grouped into three cases. In accordance with 
Fig.s \ref{fig-E-omega_c=0}-\ref{fig-E-omega_0=.2} 
we assume for the lowest pair state $m_{min}\le 0$.\\
(a) If all three pairs are identical, i.e. $M_L=3\; m_{min}$,
the triagle is equilateral. This case can occur only
in quartet states (parallel spins) and these
total orbital  angular momenta are called 'magic'.\\
(b) The thick (black) lines in Fig.\ref{fig-angles} describe the
angular distortion of a WM composed of two pairs with $m_{min}=0$ and
one pair with $m=-1$.
 Qualitatively, all pictures for two pairs with angular momentum $m_{min}$
and one with $m_{min}-1$, i.e. with $M_L=3\; m_{min}-1$, agree.
Then the WM spans a triangle where two sides are equal and one side is longer
than the other two,
because $|m_{min}-1| > |m_{min}|$ for $m_{min}\le 0$.
The Pauli principle for pair states \cite{Taut-3einB} demands that
this case can occur as in quartet (non-magic angular momentum)
 as well as in doublet states.\\
(c)  Fig.\ref{fig-angles} also shows the case for 
two pairs with $m_{min}=-1$ and
 one pair with $m=0$.
All WMs with two pairs with $m_{min}<0$ and 
one pair with $m_{min}+1$ 
 providing $M_L=3\; m_{min}+1$ look similar.
One side of the triangle is shorter than the other two equal sides
because $|m_{min}+1|< |m_{min}|$ and because this case can happen only for
 $m_{min}<0$.
The Pauli principle imposes the same restrictions
 to the quantum numbers as in case (b).

It is clear  that the distortion
has to vanish in the limit $\tilde{\omega} \rightarrow 0$, because
in this limit the electrons behave like classical particles and
their localized wave functions do not overlap.
 The distortion, however,  is a quantum mechanical effect.

As can bee seen from the Pauli principle and the minimization
 of the total energy, in ground states  the three pairs are never 
all different from each other. 
This does not apply to 
 excited states where all three angular momenta can be different
leading to a completely non-symmetric WM.
The results of our theory for excited states have to be considered 
with some caution, however, 
because its applicability to excited states (in particular c.m. excitations) 
has not yet been investigated thoroughly.
But this is not the topic of this paper.

\newpage
\section{Summary}
Using a simple unitary coordinate
 transformation and the Pauli principle, we have shown 
 that a three-electron Wigner molecule 
in a quantum dot and a magnetic field can show 
a Jahn-Teller-like distortion. 
The qualitative  fact of distortion can be shown
 analytically without any computations, 
but for a quantitative estimate we have to solve a 
one-dimensional eigenvalue problem (radial pair equation) numerically.
In the ground states the Wigner molecule is either equilateral or isosceles,
whereas excited states include completely non-symmetric 
geometries as well.
If the state is a  quartet ($S=3/2$) and the orbital angular momentum
is a magic quantum number ($M_L=3\; m ; m=$ integer), the triangle is 
equilateral. In the doublet state ($S=1/2$)
and for $M_L=3 m+1$ one of the sides is
longer, and for $M_L=3 m-1$
one of the sides is shorter than the other two.

\newpage
\section{Appendix}
\subsection{ Estimate of the ratio x/R}
For ${\bf B}=0$ the ratio x/R, which is the relevant parameter 
for the decoupling in (\ref{h-trans}), can be estimated easily.
The definition of the new coordinate $\bf x$ in (\ref{diff}) reads 
\begin{equation}
{\bf x}={\bf R}-\frac{1}{\sqrt{3}}\; \Delta {\bf r}
\label{def-x}
\end{equation}
where we dropped the indexes and
denote the electron distance by $\Delta {\bf r}$.
An idea of the  e-e-distance can be obtained from the classical
e-e-distance in the ground state which is 
$\Delta r_{cl}=3^{1/3}\; \omega_0^{-2/3}$.
The width $R_0$ of the c.m. probability distribution
 can be deduced from the exactly known c.m.
wave function (see below) providing
$$R_0=\int d^2{\bf R} \; R\; |\Phi_{c.m.}({\bf R})|^2 =
(\sqrt{\pi}/6)\;\omega_0^{-1/2}$$
Formula (\ref{def-x}) with $\Delta r/\sqrt{3} \ge R$ provides 
 for the modulus
\begin{equation}
\frac{1}{\sqrt{3}}\; \frac{\Delta r_{cl}}{R_0}-1 \le \frac{x}{R} \le 
\frac{1}{\sqrt{3}}\; \frac{\Delta r_{cl}}{R_0}+1
\end{equation} 
Because 
$\Delta r_{cl}/R_0= (6\cdot 3^{1/3}/\pi^{1/2})\;
\omega_0^{-1/6} \rightarrow \infty$
 in the Wigner limit \mbox{$\omega_0 \rightarrow 0$},
we conclude \mbox{ $x/R \rightarrow \infty$ }
and for the small decoupling parameter $R/x \rightarrow 0$.

\subsection{Validity of the approximation {\bf X}={\bf R}=0}
\begin{figure}[h]
\begin{center}
\includegraphics[width=10cm,angle=270]{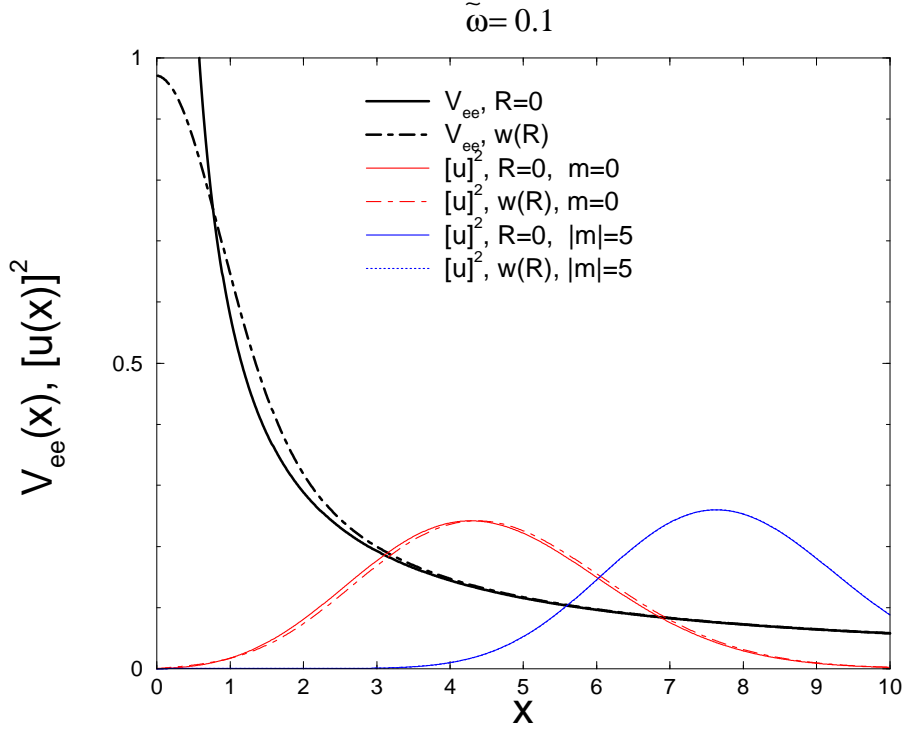}
\caption[]{(Color online) Comparison of the e-e-interaction part of the
effective pair potential for ${\bf R}=0$ (thick,full,black)
with the result using the c.m. distribution $w(R)$
 (thick,broken,black) for $\tilde{\omega}=0.1$.
The radial parts of the pair function
(\ref{rad-eq}) for $|m|=0$ and $5$ and for ${\bf R}=0$ (full) and
$w(R)$ (broken) are colored.}
\label{fig-Vee-av}
\end{center}
\end{figure}

Here we want to estimate the error  due to the neglect of 
${\bf X}={\bf R}$ in\\
i) the e-e-interaction term
in the transformed Hamiltonian (\ref{h-trans})\\
ii) the pair length considered in Sect.4.\\
In a previous paper \cite{Taut-3einB} the e-e-interaction term
was expanded in a multi-pole series (for ${\bf X}<{\bf x}_i$)
and the correction terms were considered in 
perturbation theory. 
The problem with this approach is that the resulting series converges 
slowly.
Additionally,  it provides only the total energy in a simple way. 
Therefore, we adopted here a different approach.  
Because the probability distribution $w(X)$
 of the c.m. ${\bf X}={\bf R}$ is known exactly from Kohn's theorem 
(see Appendix C) 
we averaged the corresponding quantities containing ${\bf X}={\bf R}$ 
with $w(X)$ as weight function. 
(For the estimates in this Appendix we adopted
for  the angular momentum of the c.m. system $m=0$.)\\
We replaced the e-e-interaction
 potential
\begin{equation}
V_{ee}({\bf x}_i,{\bf X})
=\frac{1}{\sqrt{3}} \; {1 \over |{\bf x}_i-{\bf X}|}
\label{Vee}
\end{equation}
by the averaged potential
\begin{equation}
 \overline{V}_{ee}(x_i)=\int d^2{\bf X} \;
w(X)\cdot V_{ee}({\bf x}_i,{\bf X})
\label{Vee-av}
\end{equation}
This virtually means using a smoothed e-e-potential
 (see Fig.\ref{fig-Vee-av}).
Unlike perturbation theory, this approach does not 
destroy the independent-pair picture.
Putting ${\bf R}={\bf X}=0$ can also be viewed as using a 
$\delta$-function-like c.m. distribution as found in
the ground state of classical mechanics.

\begin{figure}[h]
\begin{center}
\includegraphics[width=10cm,angle=270]{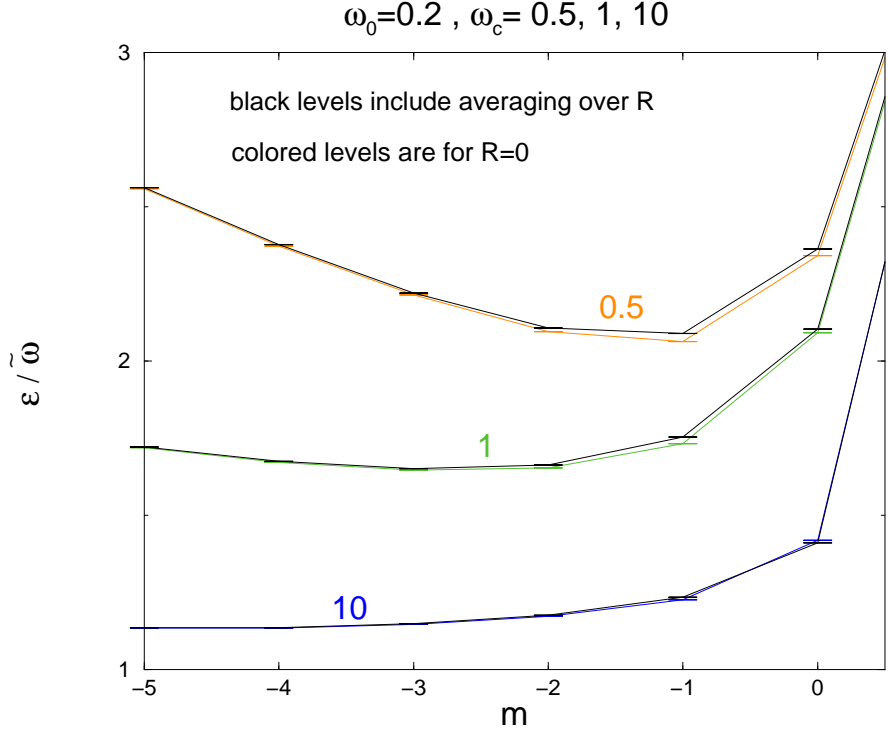}
\caption[]{(Color online) Comparison of the pair energies
(lowest state for given $m$) for
 for ${\bf R}=0$ (colored) and for the c.m. distribution $w(R)$
(black). The confinement frequency is the same as in
Fig.\ref{fig-E-omega_0=.2} and the cyclotron frequencies
 are indicated at the levels.
}
\label{fig-E-av}
\end{center}
\end{figure}

Despite the fact that the averaged e-e-potential in Fig.\ref{fig-Vee-av}
 deviates or small $x$ strongly from the curve for ${\bf R}=0$,
the change in the radial part of the pair function is very small.
This comes from the fact that the probability density $[u(x)]^2$ is small
for those $x$, where the averaged e-e-potential is changed by averaging.
For $|m|=5$ the two curves for ${\bf R}=0$ and $w(R)$ are
hardly distinguishable.
The maximum shift in the
pair energies  due to averaging
shown in Fig.\ref{fig-E-av} is only  about 1\%.

Now we are considering the pair length.
The relation between electron distances ${\bf r}_i-{\bf r}_k$, 
the new coordinates ${\bf x}_l$ and the c.m. vector ${\bf R}$ is 
given in (\ref{diff}). Denoting ${\bf r}_i-{\bf r}_k$  for a chosen pair 
by  ${\bf r}$ omitting the index at ${\bf x}_l$, we have
\begin{equation}
{\bf r}'=\frac{1}{\sqrt{3}} \;{\bf r}={\bf R}-{\bf x}
\end{equation}
Under the assumption of statistical independence,
the probability density $w_r({\bf r}')$ 
follows from the  known  probability densities 
\begin{equation}
w_x(x)=\frac{1}{2\pi} \frac{[u(x)]^2}{x}
\end{equation}
and
\begin{equation}
 w_R(R)=\frac{1}{2\pi} [{\cal R}(R)]^2
\end{equation}
using
\begin{equation}
w_r(r')=\int d^2{\bf R}\; w_R(R)\; w_x(|{\bf R}-{\bf r}'|)
\label{prob-r}
\end{equation}

\begin{figure}[h]
\begin{center}
\includegraphics[width=10cm,angle=270]{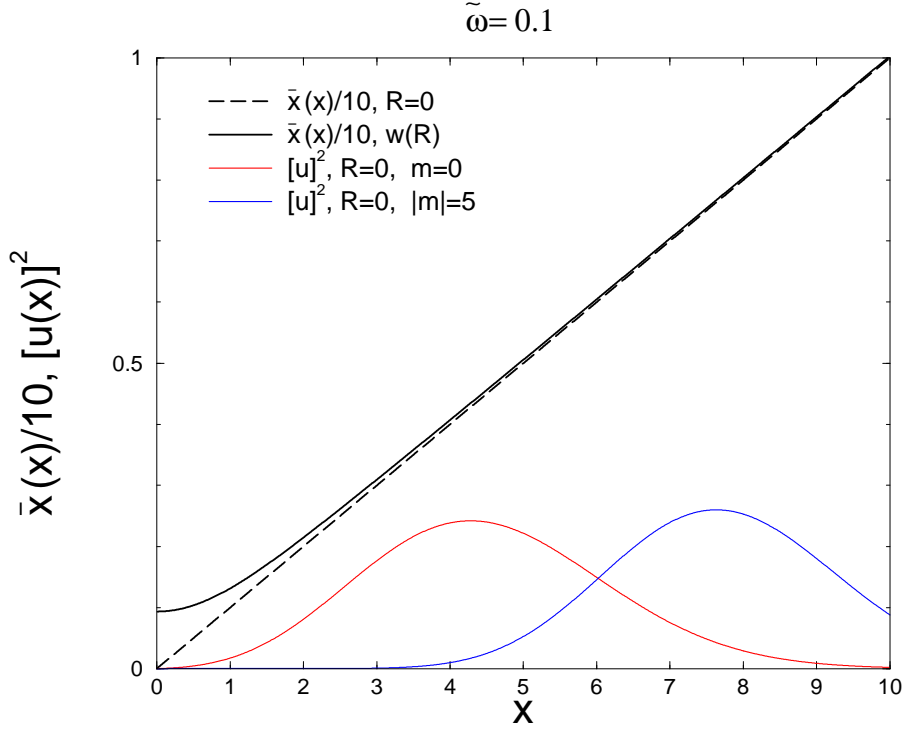}
\caption[]{(Color online) Weight function
$\bar{x}(x)$ for the calculation of the pair length 
for $R=0$ and for the finite distribution $w(R)$ (black) 
as well as the radial pair functions 
with ${\bf R}=0$ (colored) for the pair momenta $|m|=0$ and $5$.
}
\label{fig-xav}
\end{center}
\end{figure}

For the definition of the pair length we use the 
expectation value of the e-e-distance
\begin{equation}
<r'>=\int d^2{\bf r}'\; r' \; w_r(r')
\end{equation} 
Using (\ref{prob-r}), rescaling $<r>=\sqrt{3}<r'>$ and after 
changing one of the integration variables we obtain
\begin{equation}
<r>=\sqrt{3} \; \int d^2{\bf x}\; \bar{x}(x)\; w_x(x)= 
   \sqrt{3} \; \int dx \;\bar{x}(x)\; [u(x)]^2
\label{r-av-finite}
\end{equation}
with the weight function
\begin{equation}
\bar{x}(x)=\int d^2{\bf R}\; w_R(R)\; |{\bf x}-{\bf R}|
\end{equation}
For $\delta$-function-like c.m. distribution 
(corresponding to the approximation \mbox{${\bf R}=0$})
we have $\bar{x}(x)=x$ and the result from (\ref{r-av-finite}) agrees
 with the result from (\ref{x-av}).
Fig.\ref{fig-xav} shows that $\bar{x}(x)$ calculated with a finite 
distribution for ${\bf R}$  deviates only for small $x$ from  the result 
with ${\bf R}=0$. 
In this region, however, the radial pair function is small and 
the pair length for both approaches shown in 
Fig.\ref{fig-distance-diff-methods} deviate only marginally.

In the end we want to emphasize that
the main conclusions of this paper, namely the qualitative statements
about the distortion of the WM, are not influenced by averaging 
over $\bf R$.
There are only  small shifts in the pair energies and pair length.
There is a hand-waving argument for this robustness against changes 
in the c.m. vector.
It is connected to the Generalized Kohn theorem,
which states that the c.m. coordinate decouples
 exactly from properly defined relative coordinates (see Appendix C).
Therefore, no matter how large the width of the c.m. distribution is, 
it has
no influence on the corresponding relative coordinates
and on the internal structure of the WM.
However, we have to consider that 
our relative coordinates ${\bf x}_i$ are not
completely decoupled from the c.m. vector, but are coupled weakly 
as indicated by the small shifts.


\subsection{Jacobi transformation}
Applying the Jacobi transformation (see e.g. \cite{Jacak})
\begin{equation}
 \left[ \begin{array} {c} \mbox{\boldmath$ \rho$}_1\\
 \mbox{\boldmath$ \rho$}_2\\{\bf R}\end{array} \right]=
\left[ \begin{array} {ccc}1&-1&0\\
1/2& 1/2&-1\\1/3&1/3&1/3\end{array}\right]
\left[ \begin{array} {c} {\bf r}_1 \\
 {\bf r}_2 \\ {\bf r}_3 \end{array} \right]
\label{Jacobi-trafo}
\end{equation}
to (\ref{h-orig}) decouples the c.m. vector $\bf R$
from the relative coordinates $\mbox{\boldmath$ \rho$}_i$
\begin{equation}
H=H_{cm}({\bf R},{\bf P}) +H_{rel}(\mbox{\boldmath$ \rho$}_i
,\mbox{\boldmath$ \pi$}_i)
\end{equation}
where ${\bf P}=\sum_i {\bf p}_i$ is the total momentum
and $\mbox{\boldmath$ \pi$}_i$ are the canonical momenta belonging to
$\mbox{\boldmath$ \rho$}_i$.
The distance between the electrons and thus the internal structure of the WM
depends only on the relative coordinates $\mbox{\boldmath$ \rho$}_i$
\begin{eqnarray}
{\bf r}_1-{\bf r}_2&=&\mbox{\boldmath$ \rho$}_1
 \nonumber \\
{\bf r}_2-{\bf r}_3&=&-\frac{1}{2}\;\mbox{\boldmath$ \rho$}_1
+ \mbox{\boldmath$ \rho$}_2
\label{diff-Jacobi}\\
{\bf r}_3-{\bf r}_1&=&-\frac{1}{2}\;\mbox{\boldmath$ \rho$}_1
- \mbox{\boldmath$ \rho$}_2
\nonumber
\end{eqnarray}
This means that there is no correlation between the c.m. vector 
and the structure of the WM.
For the estimates in Appendix A  we need only the c.m. Hamiltonian which
reads (for arbitrary electron number $N$)
\begin{equation}
H_{cm}=\frac{1}{N} \biggl[ \;
{1\over 2}{\bf P}^2
+{1\over 2} \; \tilde\omega^2 \; (N {\bf R})^2 + {1\over 2} \;\omega_c \; (N {\bf R}) \times {\bf P}
 \;\biggr]
\label{h-cm}
\end{equation}
The c.m. eigenvalues are independent of $N$ and agree with (\ref{non-int}),
but the   eigenfunctions are homogeneously compressed by a factor of  $N$.
The latter read in polar coordinates ${\bf R}=(R,\alpha)$
\begin{equation}
\Phi_m({\bf R})=\frac{e^{im\alpha}}{\sqrt{2 \pi}} \; {\cal R}_{|m|}(R)
\label{wf-cm}
\end{equation}
with the radial part for the lowest state for given $m$
\begin{equation}
{\cal R}_{|m|}(R)=\sqrt{\frac{2}{|m|!}} \;
(\sqrt{\tilde{\omega}}N) \;
 (\sqrt{\tilde{\omega}}N R)^{|m|} \;
e^{-(\tilde{\omega}/2)(NR)^2}
\label{R-cm}
\end{equation}
With (\ref{wf-cm}) and (\ref{R-cm})   the probability density for $\bf R$ 
reads 
\begin{equation}
w_m(R)=(1/2\pi)\;[{\cal R}_{|m|}(R)]^2 .
\label{w_m(R)}
\end{equation}

{\bf Acknowledgement}
I thank H.Eschrig and M.Richter for discussions.
This work was supported by the German Research Foundation (DFG)
in the Priority Program SPP 1145.

\end{document}